\documentclass[reprint, final, aps, prb,showpacs]{revtex4-1}
\usepackage{graphicx}
\usepackage{dcolumn}
\usepackage{bm}
\usepackage{amsthm}
\usepackage{amsmath}
\usepackage{amssymb}
\usepackage{epstopdf}
\usepackage{fullpage}
\usepackage[switch]{lineno}

\begin{document}


\title{Optimization of nanocomposite materials for permanent magnets by micromagnetic simulations: effect of the intergrain exchange and the hard grains shape}

\author{Sergey~Erokhin and Dmitry~Berkov}
\address{General Numerics Research Lab, Moritz-von-Rohr-Strasse 1A, D-07749 Jena, Germany}

\begin{abstract}
In this paper we perform the detailed numerical analysis of remagnetization processes in nanocomposite magnetic materials consisting of magnetically hard grains (i.e. grains made of a material with a high magnetocrystalline anisotropy) embedded into a magnetically soft  phase. Such materials are widely used for the production of permanent magnets, because they  combine the high remanence with the large coercivity. We perform simulations of nanocomposites with Sr-ferrite as the hard phase and Fe or Ni as the soft phase, concentrating our efforts on analyzing the effects of ({\it i}) the imperfect intergrain exchange and ({\it ii}) the non-spherical shape of hard grains. We demonstrate that - in contrast to the common belief - the maximal energy product is achieved not for systems with the perfect intergrain exchange, but for materials where this exchange is substantially weakened. We also show that the main parameters of the hysteresis loop - remanence, coercivity and the energy product - exhibit non-trivial dependencies on the shape of hard grains, and provide detailed explanations for our results. Simulation predictions obtained in this work open new ways for the optimization of materials for permanent magnets.
\end{abstract}


\maketitle




\section{Introduction} 
\label{Intro}

Magnetic materials which can be used for the manufacturing of permanent magnets belong to the key materials in many high-technology applications today \cite{Stamps2014}. Modern electromotors, actuators and magnetic-field-based sensors  (to name only a few applications) require high-performance magnets. The effectiveness of these magnets is usually estimated by the value of the maximal  energy product $(BH)_{\mathrm{max}}$ achieved in the second quadrant of their $B-H$ hysteresis loop. One of the most promising ways to obtain large values of this  parameter is the usage of magnetic nanocomposites, \textit{i.e.} materials which combine a high coercivity of a magnetically hard phase (the phase made of a material possessing a large magnetocrystalline anisotropy) with the high  saturation magnetization of the another (soft) phase. 

At present, best performance magnets are produced basing on rare-earth metals (NdFeB, SmCo) and employing the precise fabrication control (see, e.g. \cite{Gutfleisch2000}). Unfortunately, these magnets are relatively expensive and subject to the availability and price fluctuations due to the high volatility of the rare earth elements market.

Another very important class of materials for permanent magnets is represented by nanocomposites containing ferrites as the hard phase. The energy product of Co-, Ba- and Sr-ferrite-based magnets is sufficient for many applications of permanent magnets, e.g. in microwave devices, telecommunication, recording media, and electronic industry. Another important advantage of these materials have a much better temperature and corrosion resistance than NdFeB-based magnets (see, e.g. Ch. 12.2 in \cite{Coey2010magnetism}).  In addition, situation with the production of ferrite-based magnets is more stable, and costs are much lower due to the wider availability of the corresponding raw materials.

The specified performance of a nanocomposite material can in principle be achieved by tailoring  various parameters of a nanocomposite, such as relative fractions of the soft and hard phases, size and shape of hard grains, mutual arrangement of grains belonging to different phases, and the quality of the intergrain boundaries. The development of new magnets of any type requires the thorough understanding of the relationship between their microstructure and magnetic properties \cite{Skomski2013}. Advanced structural experimental techniques can provide a very important information; well known examples are, e.g. the electron Bragg scattering diffraction studies of the grain alignment in sintered NdFeB \cite{Matsuura2013} or the X-ray diffraction along with the high-resolution transmission electron microscopy applied for the measurements of the grain size distribution in soft magnetic alloys \cite{Groessinger2005}.

Recent publications of different scientific collaborations have shown that in order to study the question of the microstructure-magnetism relationship in details, it is absolutely necessary to employ numerical modeling - in particular, micromagnetic simulations, combining it with other experimental methods like  3D atome probe \cite{Liu2013}, energy-dispersive X-ray spectroscopy \cite{Sepehri-Amin2013} or magnetic neutron scattering \cite{Loffler2005, Ogrin2006}. The usage of micromagnetic modeling in the development process allows the {\it a priori} performance optimization of permanent magnets, by predicting magnetic characteristics of a nanocomposite material before its actual manufacturing. 

In general, micromagnetic simulations are perfectly suited for modeling of magnetic composites, because the typical micromagnetic characteristic length - a few nanometers - allows to resolve very well the magnetization distribution inside the nanocomposite grains with sizes of several tens of nanometers. However, corresponding simulations require an enormous computational effort due to two main reasons. First, a very fine mesh of finite elements is required for the adequate approximation of each single grain as a geometrical object with a complicated shape. Second, a large number of soft and (especially) hard grains should be present in the simulated volume, in order to study the magnetization reversal as a collective phenomenon and to obtain a sufficiently accurate statistics for these disordered systems (see a detailed discussion of these issues in \cite{Erokhin2012prb1,Erokhin2012prb2,Michels2014jmmm}).

In recent years, major simulation effort was devoted to the understanding of rare-earth-based materials, with the strategic goal  "to push the border" of the conventional (non-superconducting) magnets. In the first line, a large amount of numerical research was carried out for composites based on NdFeB \cite{Gao2003, Liu2013, Sepehri-Amin2013, Saiden2014, Yi2016} and PrFeB \cite{Rong2006, Zheng2008, He2012}; some results have been published also for SmCo and similar compounds \cite{Chen2006}. 

For NdFeB-based materials, the question of the internal magnetization structure (mainly vortex) in a single grain have been studied \cite{Gao2003}. The influence of the soft phase concentration (Fe or FeB), the hard grain size \cite{Saiden2014} and (very recently) magnetic behaviour of hard grains for several different shapes \cite{Yi2016} was investigated. Special attention was paid to Nd-rich compositions, where it has been shown that the enrichment with Nd leads to the coercivity enhancement due to the concentration of the additional Nd on intergrain boundaries \cite{Sepehri-Amin2013,Liu2013}.

Studies of composites based on PrFeB and containing Fe as the soft phase have been devoted to the increasing role of the magnetodipolar interaction with the growing soft phase fraction \cite{Rong2006}, correlation of the magnetization reversal of soft and hard grains \cite{Zheng2008} and the effect of the hard grains alignment \cite{He2012}. For SmCo-like materials different mechanisms of the magnetization reversal when changing the angle between the anisotropy axis and the applied field were identified \cite{Chen2006}.

Somewhat apart from the main road lie the simulations of a highly interesting yet not really applicable class of MnBi-based materials. Here the influence of the soft phase concentration (Co and FeCo) and the orientation degree of the hard grain anisotropy axes was studied numerically in \cite{Li2015}.

In contrast to the rare-earth-based composites, materials based on various ferrites as the hard phase have not been studied - up to our knowledge - by micromagnetic simulations, although this class of materials becomes increasingly important for reasons listed above.  In the present paper we address this challenge, starting with detailed studies of the effect of the intergrain exchange coupling and the influence of the hard grain shape on the material properties in nanocomposites $\mathrm{SrFe_{12}O_{19}/Fe}$ and $\mathrm{SrFe_{12}O_{19}/Ni}$. 

Our polyhedron-based micromagnetic algorithm provides a high statistical accuracy of simulated results, because we are able to handle systems containing a few thousand grains, including the ability to resolve a possibly non-trivial magnetization distribution in every grain. Due to the flexibility of our mesh generation method, a nearly arbitrary grain shape can be adequately approximated, so that a simulated sample may include the grains of different shapes and sizes. 

The paper is organized as follows: in Sec. \ref{AlgorGen} we explain the mesh generation method which we use to create a polyhedron mesh for a system containing non-spherical hard grains embedded into a magnetically soft matrix. Evaluation of micromagnetic energy contributions in our methodology is also briefly presented.  Sec. \ref{Results} contains simulation results. Here, subsection \ref{ExchWeakEffect} is devoted to the analysis of the influence of the intergrain exchange weakening on the magnetization reversal, whereas subsection \ref{GrainShapeEffect} deals with the effect of non-spherical shapes of the hard grains. Both subsections contain a detailed physical discussion of results obtained. We conclude with the summary of our findings and possible perspectives for the improvement of ferrite-based composites in Sec. \ref{Conclusion}.

\section{Our micromagnetic methodology for simulation of nanocomposites}
\label{AlgorGen}
To overcome the difficulties in modeling magnetic nanocomposites using standard micromagnetic methods (finite difference and tetrahedral finite elements), we have developed \cite{Michels2014jmmm} a novel micromagnetic methodology based on a discretization of magnetic materials using polyhedrons of a special type. This methodology combines the flexibility of general finite element schemes for the geometrical description of a nanocomposite structure with the possibility to use Fast Fourier Transformation for the calculation of the most time consuming contribution to the total energy - magnetodipolar interaction.

\subsection{Mesh generation for grains of a general shape}
\label{MeshGen}

\begin{figure}[h!]
\centering
\resizebox{0.95\columnwidth}{!}{\includegraphics{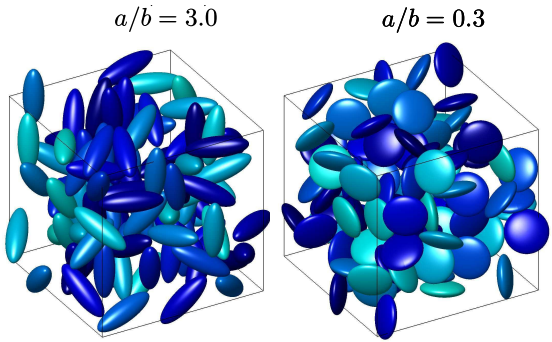}}
\caption{(color online) Examples of the spatial distribution of hard crystallites (soft crystallites not shown) in simulated samples for different aspect ratio $a/b$ of corresponding ellipsoids of revolution (see text for details).}
\label{fighard113_311}
\end{figure}

\begin{figure}[h!]
\centering
\resizebox{0.6\columnwidth}{!}{\includegraphics{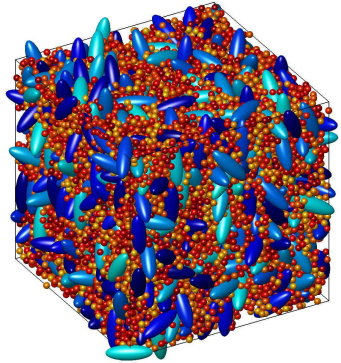}}
\caption{(color online) Example of a microstructure (hard and soft phases) used in our modeling of nanocomposites.}
\label{figellhardsoft}
\end{figure}

One of the main questions discussed in this paper is the influence of the non-spherical (spheroidal) shape of hard grains on the magnetic behavior of nanocomposites. For such a system, we had to introduce two additional steps into the mesh generation procedure described in our previous publications \cite{Erokhin2012prb1,Erokhin2012prb2,Michels2014jmmm}. Namely, in the present work the mesh for hard grains  is generated separately from the soft phase mesh (Fig.\ref{fighard113_311}) (additional step 1), and then both systems are merged (Fig. \ref{figellhardsoft}) (additional step 2). 

As in our standard methodology, we start from the generation of mesh consisting of small nearly spherical polyhedrons with the sizes less the characteristic micromagnetic length. These polyhedrons will be used in micromagnetic simulations as corresponding finite elements (see \cite{Erokhin2012prb1,Erokhin2012prb2,Michels2014jmmm} for details).

Next, we generate a system of non-overlapping ellipsoidal particles (additional step 1) with sizes and shapes corresponding to the hard grains of our composite. We point out that the generation of an ensemble of non-overlapping ellipsoids is computationally challenging: the evaluation of an overlapping of two ellipsoids and the introduction of the suitable overlapping criterion require a development of a special numerical scheme. We use a method described by Donev et al. \cite{Donev2005}, which is based on the Perram-Wertheim overlap potential and provides a suitable parameter describing the overlapping degree of two ellipsoids. This parameter is then used in the model of interacting particles with a short-range repulsive potential, where initially ellipsoids are placed randomly, but due to the nature of this potential the number of overlaps is continuously decreasing.

At the second additional step, these ellipsoids are mapped on the system of (much smaller) polyhedrons used as mesh elements in our micromagnetic simulation. By this mapping all mesh elements with centers inside ellipsoids, are assigned  to the hard phase and the rest of elements - to the soft phase. All mesh elements belonging to the same hard grain have the same direction of the anisotropy axes; this direction coincides with the rotational symmetry axis of the ellipsoid. Note, that the discretization of crystallites of the soft and hard phases is based on mesh elements of the same size.

\subsection{Energy contributions and minimization procedure}
\label{EnergyCalc}

In our simulations we take into account all four standard contributions to the total magnetic free energy: energy in the external field, magnetocrystalline anisotropy energy, exchange stiffness and magnetodipolar interaction energies \cite{Erokhin2012prb1,Erokhin2012prb2,Michels2014jmmm}. 

The local energy parts - energy in the external field and the magnetocrystalline anisotropy energy - are computed in a standard way, multiplying the corresponding energy densities by the volume of finite elements (polyhedrons in our case) and summing over all these elements. The magnetodipolar field and energy are computed using the optimized version of the lattice Ewald method for disordered systems. In this algorithm, the mapping of the initial (disordered) system of mesh elements on the translationally invariant regular lattice allows to keep the high speed of the lattice method (fast Fourier transformation), at the same time making the mapping errors negligibly small.

In  this work, we concentrate ourselves in particular on the effect of the intergrain exchange. Hence we remind that the exchange energy in our methodology is computed in the nearest neighbouring approximation as 
\begin{equation}
\label{EqExchEn}
E_{\rm exch} = - \frac{1}{2} \sum_{i = 1}^N \sum_{j \subset \mathrm{n.n.}(i)} \frac{2A_{ij} \, \overline V_{ij}}{\Delta r_{ij}^2} \left( \mathbf{m}_i \, \mathbf{m}_j \right) ,
\end{equation}
where $\overline V_{ij} = (V_i + V_j)/2$, $\Delta r_{ij}$ is the distance between the centers of $i$-th and $j$-th finite elements with volumes $V_i$ and $V_j$. The exchange constant $A_{ij}$ for the homogeneous bulk material is equal to the corresponding exchange stiffness constant $A$, but is obviously site-dependent in composite materials \cite{Erokhin2012prb2,Michels2014jmmm}.

For the minimization of the total magnetic energy, obtained as the sum of all four contributions described above, we use the simplified version of a gradient method employing the dissipation part of the Landau-Lifshitz equation of motion for magnetic moments \cite{Landau35, KronParkinHandbook07}. The minimization is considered as converged, when the condition for the local torque
$\max_{\{i\}}|[\mathbf m_i \times \mathbf h_i^{\rm eff}]|< \varepsilon$ is fulfilled (here $\mathbf m_i$ is a normalized magnetic moment of the $i$-th mesh element and $\mathbf h_i^{\rm eff}$ is the corresponding effective field; the value $\varepsilon = 10^{-3}$ was found to be small enough for our quasistatic minimization procedure).

Further details of our method can be found in \cite{Michels2014jmmm}.

\section{Results and discussion}
\label{Results}

Due to the high performance of our methodology, we are able to simulate bulk nanocomposites with ‘hard’ grains of any prescribed shape, whereby the simulated system may contain up to 600 hard grains with an average discretization of 300 mesh elements pro grain. Employing this algorithm, we could obtain systematic results with the high statistical accuracy for two model systems: $\mathrm{SrFe_{12}O_{19}/Fe}$ and $\mathrm{SrFe_{12}O_{19}/Ni}$.

For simulations of these nanocomposites we have used standard magnetic parameters of corresponding materials (see, e.g.\cite{Coey2010magnetism}, Chap. 5.3 and 11.6) which are summarized in the table below:\\

\begin{tabular}{|c|c|c|c|}
\hline 
  &   $\mathrm{SrFe_{12}O_{19}}$  & $\mathrm{Fe}$ &  $\mathrm{Ni}$  \\ 
\hline 
$M_s$(G) & 400 & 1700 & 490 \\ 
\hline 
Anis. kind & uniaxial & cubic & cubic \\ 
\hline 
$K$ (erg/cm$^3$) & $4.0 \cdot 10^6$  & $5.0 \cdot 10^5$ &  $ -4.5 \cdot 10^4$ \\ 
\hline 
$A$ (erg/cm) & $0.6 \cdot 10^{-6}$ & $2.0 \cdot 10^{-6}$ & $0.8 \cdot 10^{-6}$ \\ 
\hline 
\end{tabular}

\vspace{5mm}
The  average grain volume for all phases was chosen to be equal to the volume of a spherical grain with the diameter $D = 25 \rm{\,nm}$. The volume concentration of the hard phase in all presented simulations was $c_{\rm{hard}} = 40 \, \%$.

\subsection{Effect of the exchange weakening in $\mathrm{SrFe_{12}O_{19}/Fe}$}
\label{ExchWeakEffect}

One of the central questions for permanent magnets made of nanocomposite materials is the dependence of magnetic properties on the exchange weakening between different grains. This weakening is unavoidable in real systems, because it is nearly impossible to obtain perfect intergrain boundaries. The quality of these boundaries strongly depends on the concrete method used for the manufacturing of a nanocomposite and substantial efforts has been devoted to obtaining materials with more perfect intergrain boundaries (and especially boundaries between grains belonging to different phases) in order to achieve better exchange coupling. However, recently  \cite{QuesadaAdvElMat_2016} is was demonstrated both experimentally and theoretically that the perfect intergrain exchange may strongly decrease the performance of a magnetic nanocomposite material, so that this question requires a detailed theoretical study.

The exchange weakening in our methodology is defined by multiplying the exchange energy of neighboring mesh elements belonging to {\it different} crystallites by a factor $0 \leq \kappa \leq 1$. This exchange weakening coefficient is introduced into the expression (\ref{EqExchEn}) for the exchange energy as 

\begin{equation}
\label{EqExchEn_mod}
E_{\rm exch} = - \frac{1}{2} \sum_{i = 1}^N \sum_{j \subset \mathrm{n.n.}(i)} \kappa \frac{2A_{ij} \, \overline V_{ij}}{\Delta r_{ij}^2} \left( \mathbf{m}_i \, \mathbf{m}_j \right) ,
\end{equation}
if neighboring magnetic moments $i$ and $j$ are located in different grains (crystallites). From (\ref{EqExchEn_mod}) it can be seen that  $\kappa = 1$ corresponds to the perfect intergrain exchange (equal to the exchange within a bulk material) and $\kappa = 0$ means no exchange interaction at all between  different grains. 

Dependence of magnetic properties on this exchange weakening was studied for the composite $\rm{SrFe_{12}O_{19}/Fe}$ with approximately spherical hard grains (obtained from the random placement of spheres with $D = 25 \rm{\,nm}$, see Sec. \ref{MeshGen}).

\begin{figure}[h!]
\centering
\resizebox{0.95\columnwidth}{!}{\includegraphics{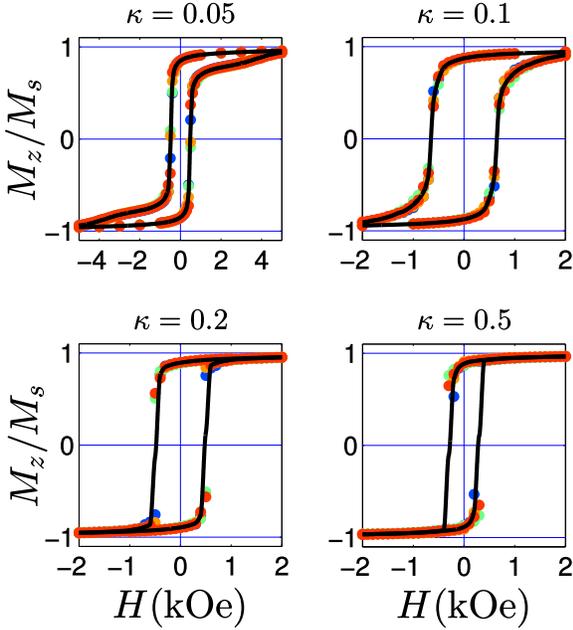}}
\caption{(color online). Simulated hysteresis curves of the nanocomposite $\mathrm{SrFe_{12}O_{19}/Fe}$ with spherical hard grains for different exchange weakening constant.}
\label{fighystFe111}
\end{figure}

The overall trend is shown in Fig. \ref{fighystFe111}, where  the evolution of hysteresis curves by increasing the exchange coupling ($\kappa = 0.0 \rightarrow 0.5$) between grains is demonstrated. Systems without ($\kappa = 0.0$) or with a strongly reduced ($\kappa = 0.05$) exchange coupling between grains exhibit the two-steps magnetization reversal. The first step - large jump on the hysteresis loop in small negative fields (see the panel for $\kappa = 0.05$ in Fig. \ref{fighystFe111}) - represents the magnetization reversal of the soft phase, which volume fraction is relatively high. The second step - reversal of hard grains in much higher fields - leads to the closure of the loop. Reversal of the hard phase occurs in fields $\sim H_K^{\rm hard}$, where the anisotropy field is defined as $H_K = 2K/M_s = \beta M_s$ ($\beta = 2K/M_s^2$ denotes the reduced anisotropy constant). The large magnitude of the magnetization jump during the first reversal step is due to the dominating contribution of the soft phase to the system magnetization: 
$m_{\rm{soft}} = c_{\rm{soft}}  \cdot M_{\rm{soft}}/(c_{\rm{hard}}  \cdot M_{\rm{hard}} + c_{\rm{soft}}  \cdot M_{\rm{soft}}) \approx 0.86$.  

For the detailed analysis of the magnetization reversal in a magnetic composite it is very useful to plot hysteresis loops for the soft and hard phase separately. Such loops can be easily obtained from simulated magnetization configurations by summing up contributions from finite elements belonging to either soft or hard phase and calculating corresponding total magnetizations of these phases.

\begin{figure}[h!]
\centering
\resizebox{0.75\columnwidth}{!}{\includegraphics{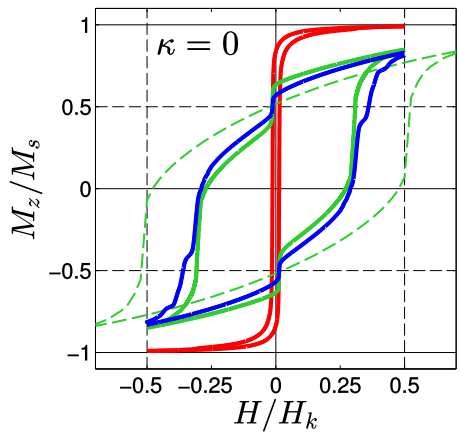}}
\caption{(color online). Simulated hysteresis loops for $\mathrm{SrFe_{12}O_{19}/Fe}$ (with spherical hard grains) without the intergrain exchange ($\kappa = 0$) presented for hard (solid blue line) and soft (solid red line) phases separately. Dashed line represents the unsheared loop of the SW model with particle parameters as for $\mathrm {SrFe_{12}O_{19}}$, solid green line - the SW loop sheared according the averaged internal field (see text for details). External field is normalized by the anisotropy field of the hard phase $H_K = \beta_{\rm h} M_{\rm h} = 20 \, {\rm kOe}$.}
\label{fighystFe111_hs}
\end{figure}

We begin our consideration from systems with an absent or very low intergrain exchange coupling, where  the dominant interaction is the magnetodipolar one. To clarify the effect of this interaction, the above mentioned magnetization reversal curves of hard and soft phases are plotted in Fig. \ref{fighystFe111_hs} for the composite without any intergrain exchange coupling ($\kappa = 0$). 

In order to understand the hysteretic behavior of both phases, it is useful to calculate the reduced anisotropy constant $\beta = 2 K/M_{\rm s}^2$, which magnitude gives (roughly speaking) the relation of the magnetocrystalline  anisotropy field to  the magnetodipolar field from the nearest neighbor in the system of spherical particles. Substitution of magnetic parameters  of our materials (see the table above) results in the values $\beta_{\rm h} = 50 (\gg 1)$ for the hard phase ($\rm{SrFe_{12}O_{19}}$) and $\beta_{\rm s} = 0.34 (\sim 1)$ for the soft phase (Fe). We note that the much higher value of $\beta$ for the hard phase is due not only to its large anisotropy constant $K$ (which is 'only' 8 times larger than by Fe), but mainly due to the much higher value of the soft phase magnetization $M_{\rm Fe}/M_{\rm SrFeO} = 4.25$, which gives an additional factor of  $\approx 18$.

Considering the magnetization reversal of the {\it soft} phase first, we note that this phase would exhibit in the absence of the magnetodipolar interaction the 'ideal' hysteresis loop for a system of non-interacting particles with the cubic anisotropy constant $K_{\rm cub} > 0$ (as it is the case for ${\rm Fe}$). Such a loop has the remanence $j_R^{\rm (0)} \approx 0.83$ and the coercivity $H_c^{\rm (0)} \approx 0.33 H_K = 0.33 \beta M_{\rm s} \approx 195 {\rm \, Oe}$ (see, e.g. \cite{Usov1997}).  The relatively low value of the reduced anisotropy for our soft phase $\beta_{\rm s} ({\rm Fe}) = 0.34$ means that the magnetodipolar interaction can considerably modify the corresponding 'ideal' hysteresis. This influence manifests itself primarily in the smoothing of the 'ideal' loop \cite{Usov1997}, as it can be seen in Fig. \ref{fighystFe111_hs}, where the loop for the soft phase of our system is shown in red. The remanence $j_R \approx 0.836$ is nearly the same and the coercivity $H_c \approx 260 {\rm \, Oe}$  increased by $\approx 30 \%$ compared to the non-interacting case. 

Unfortunately, we are not aware of any systematic theoretical studies of the magnetodipolar interaction effects in systems of 'cubic' particles, except for the paper \cite{Garcia2000}, where only simulation results for the Henkel plots are shown; any quantitative comparison with a detailed study of these effects for 'uniaxial' particles presented in \cite{Berkov1996} is meaningless due to very different energy landscapes for these two anisotropy types. For this reason, we can only suggest that the nearly unchanged remanence (compared to the 'ideal' system) is due to the interplay of the magnetodipolar interactions within the soft phase and between the soft and hard phases. The increase of $H_c$ is most probably due to the 'supporting' action of the magnetodipolar field from the hard phase onto the soft grains. Magnetization of the hard phase in our system is rather low, so that the corresponding effect is relatively small.

The non-interacting {\it hard} phase consisting of grains with the uniaxial anisotropy (as for $\mathrm {SrFe_{12}O_{19}}$) would reverse according to the ideal Stoner-Wohlfarth (SW) loop \cite{Stoner1948} with $j_R = 0.5$ and $H_c \approx 0.48 H_K \approx 10 \, {\rm kOe}$ shown in Fig. \ref{fighystFe111_hs} with the thin dashed green line. The very large value of the reduced single-grain anisotropy  $\beta_{\rm h} ({\rm SrFeO}) = 50$ for this phase means that intergrain correlations of hard phase magnetic moments are negligible. However, in our composite material hard grains are 'embedded' into the soft phase. Hence, in order to properly compare (at least in the mean-field approximation) the simulated hard phase loop - blue solid line in Fig. \ref{fighystFe111_hs} - with the SW model, we have to take into account the average magnetodipolar field $\langle H_{{\rm md}, z} \rangle = (4\pi/3) \langle M_z^{\rm soft} \rangle$ acting on a spherical particle inside a continuous medium with the average magnetization of the soft phase $\langle M_z^{\rm soft}\rangle$. 

Correction of the SW loop using this internal field (which depends on the external field via the corresponding dependence $\langle M_z (H_z)\rangle$) leads to the loop shown with the thick solid green line in Fig. \ref{fighystFe111_hs}. It can be seen that this corrected SW loop is in a good agreement with the simulated hard phase loop. Remaining discrepancies are due to local internal field fluctuations (always present in disordered magnetic systems) which are especially pronounced in our composite due to the high difference between the magnetizations of  soft and hard phases.

This analysis reveals that the first jump on the hard phase loop  in small negative fields is due to the abrupt change in the internal averaged dipolar field due to the magnetization reversal of the soft phase. The second jump - for $H_z/H_k \approx -0.3$ - is the manifestation of the singular behavior of the SW loop of the hard phase itself, which occurs for the unsheared loop at $H_{\rm cr} = -H_k/2$ (near this field $M_z \sim \sqrt{-(H_z-H_{\rm cr})}$ for $H_z < H_{\rm cr}$ \cite{Berkov1988}).

\begin{figure}[h!]
\centering
\resizebox{0.95\columnwidth}{!}{\includegraphics{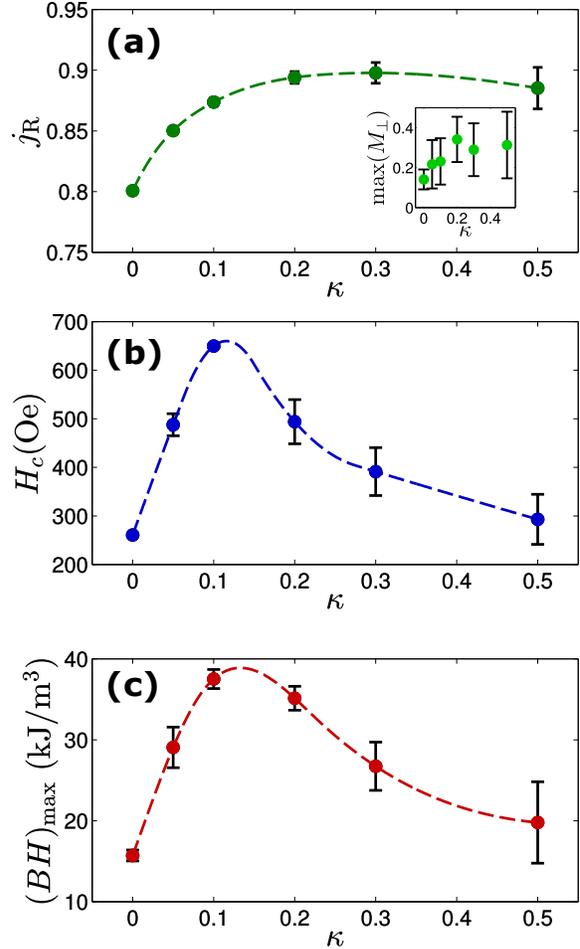}}
\caption{(color online). \textbf{(a)} Remanence, \textbf{(b)} coercivity and \textbf{(c)} energy product of simulated nanocomposite $\mathrm{SrFe_{12}O_{19}/Fe}$ with spherical hard grains as a functions of exchange weakening on the grain boundaries. Inset in (a) represents the maximal value of perpendicular (to the directions of applied field) component of magnetization during the remagnetization process. Dashed lines are paths for the eye.}
\label{figellFe111kappa}
\end{figure}

In summary, despite a relatively high saturation magnetization $M_s = 1180 \, {\rm G}$, the corresponding composite without any intergrain exchange coupling would have only a relatively small maximal energy product of $\approx 15 \, {\rm kJ/m^3}$ (see Fig. \ref{figellFe111kappa}b). The reason is its very small coercivity $H_{\rm c} \approx 250 \, {\rm Oe}$, which is determined entirely by the magnetization reversal of the soft phase in small negative fields.

Before we proceed with the analysis of the effect of the intergrain exchange coupling on the hysteretic properties of a nanocomposite, an important methodical issue should be clarified. Namely, we have to determine the maximal value of the exchange coupling (maximal value of $\kappa$), for which our simulations can produce meaningful results. 

The problem is that with increasing the coupling strength, the interaction between the grains increases, so that grains are starting to form clusters, inside which magnetic moments of constituting grains reverse  nearly coherently. The average size of such a cluster $\langle d_{\rm cl} \rangle$ obviously growths with increasing $\kappa$.  In order to obtain statistically significant results, we have to assure that $\langle d_{\rm cl} \rangle$ is significantly less (ideally much less) than the maximal system size accessible for simulations. Otherwise we might end up with the case where we are simulating the magnetization reversal of a system consisting of a single (or very few) cluster(s), so that corresponding results will be non-representative for the analysis of real experiments.  

The best quantitative method to  determine $\langle d_{\rm cl} \rangle$ is the calculation of the spatial correlation function of magnetization components perpendicular to the applied field (in our case $M_x$ and $M_y$): the average value of these components should be zero, and the decay length of their correlation functions $C_x({\bf r}) =  \langle M_x(0) M_x({\bf r}) \rangle$ (the same for $M_y$) would provide a most reliable estimation of $\langle d_{\rm cl} \rangle$. 

However, taking into account a complex 3D character of $C_{x,y}({\bf r})$, we have adopted another criterion to determine the approximate number of independent clusters contained in our simulated system. Namely, as the figure of merit we have employed the maximal value of the perpendicular component of the total system magnetization $m_{\perp} = \sqrt{M_x^2+M_y^2}/M_s$  during the magnetization reversal. 

If the system contains only one (or very few) cluster(s), than for some field during the reversal process this component should be large (close to 1), because one cluster reverses nearly in the same fashion as a single particle, i.e., its magnetization rotates as a whole without significantly changing its magnitude. Hence at some reversal stage $m_{\perp}$ would unavoidably become relatively large. In the opposite case, where a system contains many nearly independent clusters ($N_{\rm cl} \gg 1$), their components $M_{x,i}$ and $M_{y,i}$ ($i = 1,...,N$), being independent variables with zero mean, would average themselves out, leading to small values of $m_{\perp}$. 

A simple statistical analysis based on the assumption of the independence of different clusters shows that the number of such clusters can be estimated as $N_{\rm cl} \geq 1/m_{\perp}^2$. This means that up to $m_{\perp}  \approx 0.3$ we produce statistically significant results, because in this case $N_{\rm cl} \geq 10$. Corresponding analysis shows that for our  systems (containing about $\sim 5 \cdot 10^5$ finite elements) this is the case up to $\kappa \approx 0.5$, so below we show results only in this range of exchange couplings.

Simulation results showing basic characteristics of the hysteresis loop - remanence $j_R$, coercivity $H_c$ and energy product $E_{\rm max} = (BH)_{\rm max}$ - for the $\rm{SrFe_{12}O_{19}/Fe}$ composite as functions of the exchange weakening $\kappa$ are presented in Fig. \ref{figellFe111kappa}. We remind that for these simulations approximately spherical hard grains were used.

From Fig. \ref{figellFe111kappa}  it can be clearly seen that the remanence $j_R$ of this material depends on the intergrain exchange coupling relatively weak. The reason is that $j_R$ is very high already for the fully exchange decoupled composite ($j_R(\kappa = 0) \approx 0.8$). Such a high value, in turn, is due to the fact that the remanence is governed by the soft phase consisting of {\it cubical} grains. The remanence of the non-interacting (ideal) ensemble of such grains is $j_R^{\rm (0)} \approx 0.83$. This high remanence can not be significantly increased by the exchange interaction within the soft phase (as it is the case for the system of {\it uniaxial} particles with randomly distributed anisotropy axes, where $j_R^{\rm (0)} = 0.5$; see also \cite{Berkov_1998_PRB} for the analysis of a corresponding 2D system). Neither can this remanence be substantially decreased by the exchange coupling with hard grains, because their magnetization at $H_z = 0$ is still nearly aligned along the initial field direction due the strong magnetizing field from the Fe soft phase, (with its high magnetization $M_{Fe} = 1700$ G).

In contrast to $j_R$, the coercivity $H_c$ exhibits a pronounced maximum as the function of the exchange coupling $\kappa$, resulting in the corresponding maximum of the $\kappa$-dependence of the maximal energy product $(BH)_{\rm max}(\kappa)$. We will explain the reasons for the appearance of this maximum below, analyzing the hysteretic behavior of our nanocomposite for various $\kappa$.

\begin{figure}[h!]
\centering
\resizebox{0.95\columnwidth}{!}
{\includegraphics{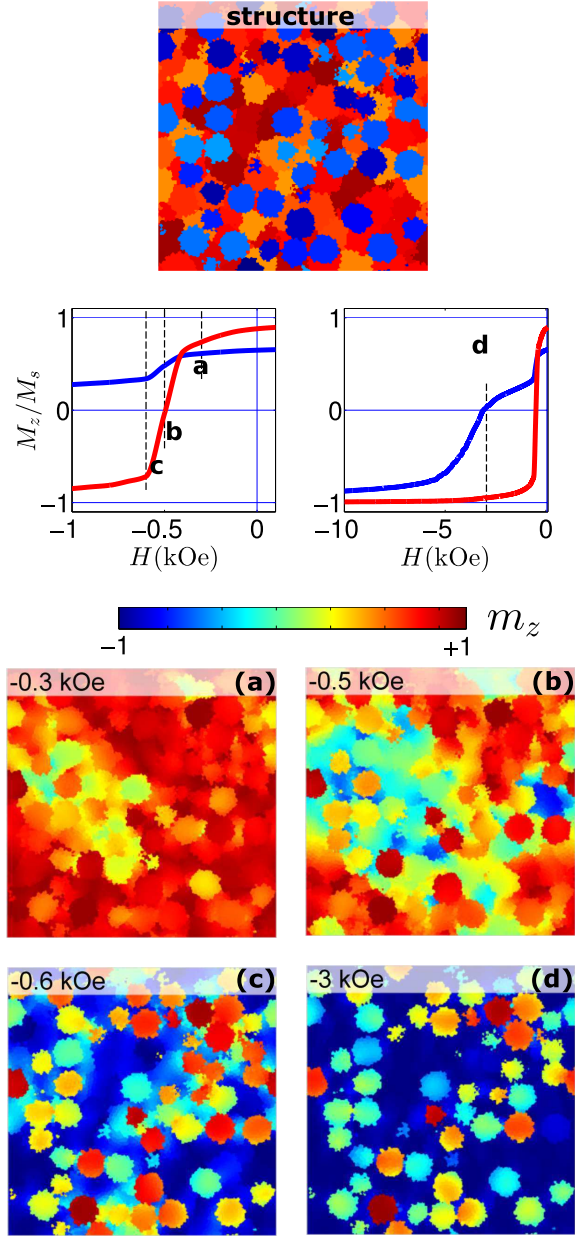}}
\caption{(color online). Magnetization reversal process for the composite with exchange weakening  $\kappa = 0.05$. From top to bottom: microstructure of the system (warm colors - soft, cold colors - hard grains); hysteresis shown as separate curves for the soft (red) and hard (blue line) phases (note different scales of the $H$-axis); magnetization configurations shown as $m_z$-maps for field values indicated on the hysteresis plots shown above.}
\label{figmagnellFe111kappa005}
\end{figure}

For the smallest non-zero $\kappa$ studied here the magnetization reversal process is visualized in Fig. \ref{figmagnellFe111kappa005}, where hysteresis loops for soft and hard phases are shown separately and the magnetization configuration  is displayed for several characteristic external fields. First, it can be clearly seen that the magnetizations of soft and hard phases reverse separately. The inspection of magnetization configurations shows that the reversal of magnetic moments starts within the soft phase (see panel (a)) around the hard grains which anisotropy axis are directed 'favorably' (i.e. deviate strongly from the initial field direction). Then the reversed area expands, occupying  even larger regions of the soft phase (panel (b)) until nearly the entire soft phase is reversed (panel (c)). Note that in the negative field corresponding to this nearly complete reversal of the soft phase, the majority of the hard phase is still magnetized approximately along the initial direction. Only in much larger negative fields (right drawing of hysteresis loops) the hard phase magnetization also starts to reverse (see panel (d)).

We emphasize here two important circumstances: although the exchange coupling between the soft and hard phases is very weak  ($\kappa = 0.05$) and the concentration of the hard phase is moderate (40\%), the 'supporting' action of the hard phase is enough to nearly double the coercivity of the soft phase and hence - of the whole system, when compared to the case of $\kappa = 0$ - see Fig. \ref{figellFe111kappa}. At the same time, due to this low exchange coupling, hard grains reverse separately from the soft phase and nearly separately from each other (see panel (c)), leading to a high coercivity of the hard phase (right drawing of hysteresis loops).

\begin{figure}[h!]
\centering
\resizebox{0.95\columnwidth}{!}
{\includegraphics{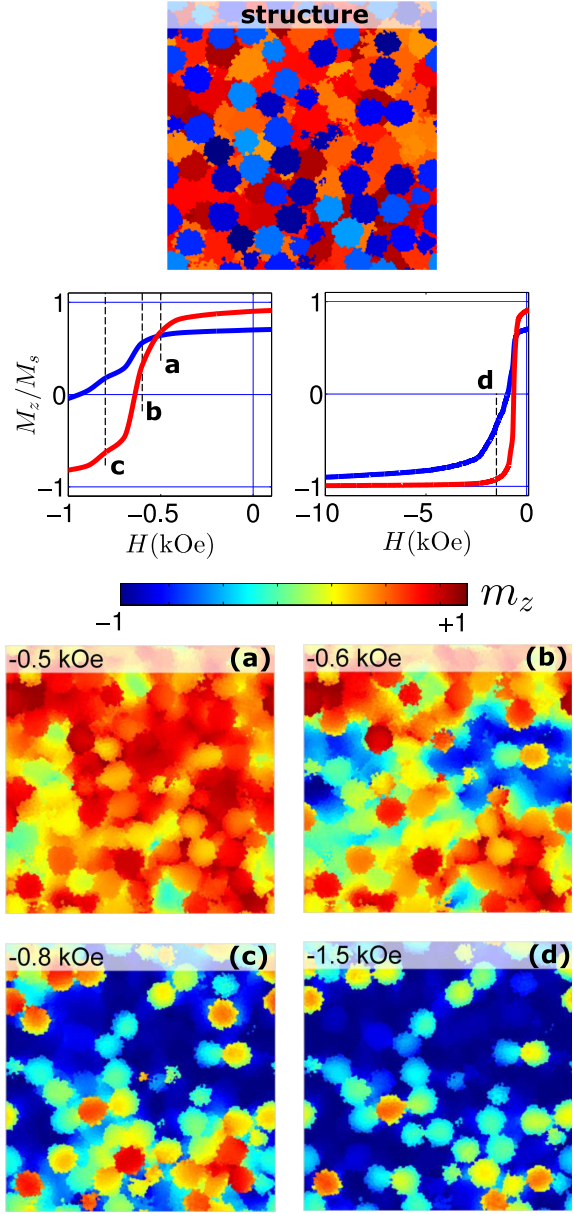}}
\caption{(color online). Magnetization reversal for the composite with the exchange weakening $\kappa = 0.10$ presented in the same manner as in Fig. \ref{figmagnellFe111kappa005}.}
\label{figmagnellFe111kappa010}
\end{figure}

For the larger exchange coupling $\kappa = 0.1$ (see Fig. \ref{figmagnellFe111kappa010}) the 'supporting' effect of the hard phase increases the coercivity of the soft phase even further (compared to $\kappa = 0.05$). At the same time, this larger coupling also leads to the much earlier reversal of the hard phase, significantly decreasing its coercivity - see hysteresis plots in Fig. \ref{figmagnellFe111kappa010}. Magnetization reversal  for this coupling starts in those system regions where the hard phase is nearly absent (due to local structural fluctuations) - see panel (b) in Fig. \ref{figmagnellFe111kappa010} - and is much more cooperative compared to the case of $\kappa = 0.05$. 

The resulting coercivity of the entire system is at its maximum, because the interphase coupling is, on the one hand, large enough to prevent the soft phase from the reversal in small fields, but on another hand, small enough to enable to the reverse of the hard phase in much higher negative fields than the soft phase.

\begin{figure}[h!]
\centering
\resizebox{0.95\columnwidth}{!}
{\includegraphics{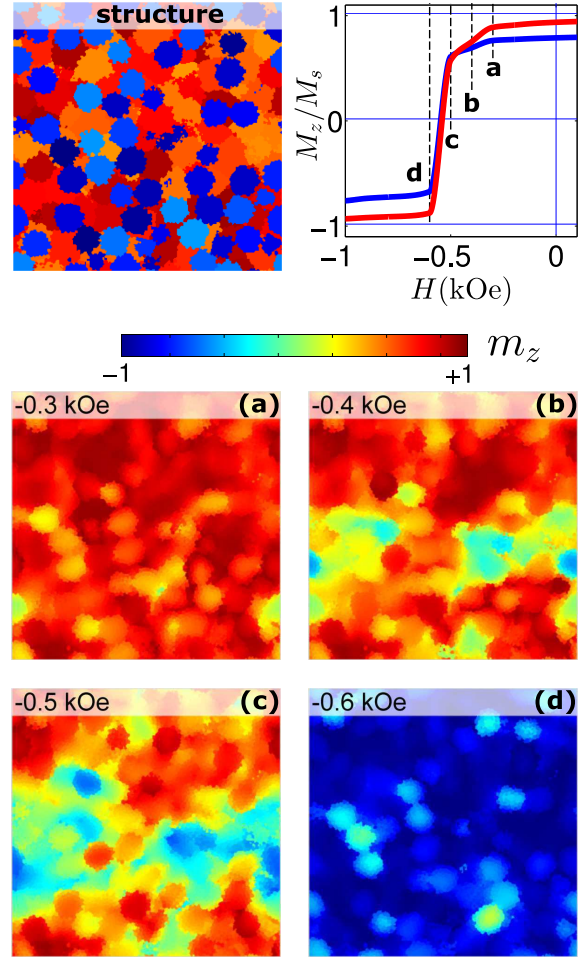}}
\caption{(color online). Magnetization reversal for the composite with the exchange weakening $\kappa = 0.20$ presented in the same way as in Fig. \ref{figmagnellFe111kappa005}. Simultaneous reversal of the hard and the soft phases is clearly visible.}
\label{figmagnellFe111kappa020}
\end{figure}

When the intergrain exchange coupling is increased further, magnetization reversal of the system becomes fully cooperative, so that the soft and hard phases reverse simultaneously (in the same negative fields) - see hysteresis loops shown in Fig. \ref{figmagnellFe111kappa020} for $\kappa = 0.2$. Spatial correlations between the microstructure and the nucleation regions for the magnetization reversal become weak, as it can be seen from microstructural and magnetic maps presented in this figure. It is also apparent that the correlation distance of the magnetization configuration strongly increases, as it was noted in the discussion above.  

The overall result is the decrease of the system coercivity, because the soft phase causes the much earlier reversal of the hard phase, so that the supporting effect of the high anisotropy of the hard phase becomes smaller. However, for this relatively low value of $\kappa=0.2$ this 'supporting' effect is still present:  $H_c(\kappa=0.2)$ is nearly twice as large as $H_c(\kappa=0)$. 

When the exchange coupling increases even further, the magnetization reversal becomes completely dominated by the soft phase due to its larger magnetization and volume fraction. In particular, for $\kappa = 0.5$ both the coercivity and the energy product are nearly the same as for $\kappa = 0$. We note that hysteresis loops for these two cases ($\kappa = 0$ and $\kappa = 0.5$) look qualitatively different, but this physically important difference (two-step vs one-step magnetization reversal) does not matter for the performance of the nanocomposite from the point of view of a material for permanent magnets.

The non-monotonous dependence of the maximal energy product on the exchange coupling $(BH)_{\rm max}(\kappa)$ can be easily deduced from the dependencies $j_R(\kappa)$ and $H_c(\kappa)$. When $\kappa$ increases from 0 to $\approx 0.1$, both remanence and coercivity increase, resulting in the rapid growth of $(BH)_{\rm max}$. For $\kappa > 0.1$, the small increase of the remanence (up to $\kappa \approx 0.2$) can not compensate the large drop of coercivity, resulting in the overall decrease of the energy product. We point out here that such a behavior occurs only when the dependence of the coercivity on the corresponding parameter (in our case the exchange weakening $\kappa$) is really strong. The case when the coercivity depends relatively weak on the parameter of interest, is analyzed in detail in the next subsection.

Summarizing this part, we have shown that, in contrast to the common belief, there exist an {\it optimal} value of the interphase exchange coupling in a soft-hard nanocomposite which provides the maximal energy product. This optimal value obviously depends on the fractions of the soft and hard phases, but it is very likely that the optimal coupling should be significantly less than the perfect coupling ($\kappa = 1$) for all reasonable compositions in this class of materials.

This important insight opens a new route for the optimization of the permanent magnet materials.

\subsection{Effect of the grain shape of the hard phase in $\mathrm{SrFe_{12}O_{19}/Fe}$ and $\mathrm{SrFe_{12}O_{19}/Ni}$ composites}
\label{GrainShapeEffect}

One of the intensively discussed questions when optimizing the nanocomposite materials for permanent magnets is whether the materials containing the hard grains with the {\it non-spherical} shape could provide an improvement of the energy product for corresponding composites (see corresponding references in the Introduction). 

The standard argument in favor of the possible improvement of $E_{\rm max}$ is the additional shape anisotropy of non-spherical particles. For an elongated (prolate) ellipsoid of revolution this anisotropy could increase the already present magnetocrystalline anisotropy (mc-anisotropy), thus enhancing the coercivity of the hard phase and hence - the energy product. Below we will demonstrate that this line of arguments is not really conclusive and that the grain shape effect may be even the opposite - the energy product can be larger for a material containing {\it oblate} hard grains.

Before proceeding with the analysis of our results, we emphasize, that the {\it relative} contribution of the shape anisotropy can be approximately the same for rare-earth and ferrite-based materials. The former materials have a much larger mc-anisotropy $K_{\rm cr}$, so that on the first glance shape effects for rare-earth 'hard' grains should be much smaller. But the the relation between the shape anisotropy and the mc-anisotropy contributions is determined not only by the value of $K_{\rm cr}$, but by the reduced anisotropy constant $\beta = 2K_{\rm cr}/M_s^2$, which gives, roughly speaking, the relation between the mc-anisotropy energy and the self-demagnetizing energy of a particle.

The presence  of the material magnetization in the denominator of the expression for $\beta$ makes this constants for both material classes very similar. For example, the mc-anisotropy $K_{\rm cr} \approx 4.6 \times 10^7 {\rm \, erg/cm^3}$ for ${\rm Nd_2 Fe_{14} B}$ is more than one order of magnitude larger than its counterpart $K_{\rm cr} \approx 4 \times 10^6 {\rm \, erg/cm^3}$ for $\rm{SrFe_{12}O_{19}}$. However, the much lower magnetization $M_s \approx 400 {\rm \,  G}$ of $\rm{SrFe_{12}O_{19}}$ compared to $M_s \approx 1300 {\rm \,  G}$ of ${\rm Nd_2 Fe_{14} B}$ makes the difference between reduced anisotropies of these materials quite small: $\beta_{\rm NdFeB} \approx 60$, whereas $\beta_{\rm SrFeO} \approx 50$. 

In the language of the anisotropy field we have to compare the values of the mc-anisotropy field $H_K = \beta M_s = 2 K_{\rm cr}/M_s$ with the values of the {\it magnetizing} magnetodipolar field, which attains its maximal value $H_{\rm dip}^{\rm max} = 2 \pi M_s$ for a needle-like particle. Corresponding relation $H_{\rm dip}^{\rm max}/H_K = \pi M_s^2/K_{\rm cr} = 2 \pi /\beta$ is $\approx 10.5$ for ${\rm Nd_2 Fe_{14} B}$  and $\approx 12.5$ for $\rm{SrFe_{12}O_{19}}$. This means that in the best case the effect of the shape anisotropy for both material classes can achieve $\approx 20\%$, what would be a non-negligible improvement on a highly competing market of modern permanent magnet materials.

Unfortunately, several circumstances are expected to strongly diminish the shape anisotropy contribution. First, the estimate above holds for a strongly elongated particle; for ellipsoidal particles with a realistic aspect ratio $a/b \sim 2 -3 $ ($a$ is the length of the axis of revolution) the shape anisotropy field is only about half its maximal value. Second, this estimation holds for a single-domain particle, whereas strongly elongated or nearly flat particles acquire a multi-domain state much easier than the spherical ones, because the domain wall energy for strongly non-spherical particles is much smaller, than for a sphere. Finally, the relation derived above is true only for an isolated particle, and hard grains in nanocomposites are always embedded into a soft phase or are in a close contact with another hard grains.

For these reasons we have performed a detailed numerical study of the dependence of hysteresis properties on the hard grain shape for nanocomposite $\rm{SrFe_{12}O_{19}/Fe}$ and - for comparison - for $\rm{SrFe_{12}O_{19}/Ni}$ . For this purpose we have simulated magnetization reversal in these composites with the hard grains having the shape of ellipsoids of revolution (spheroids) with the aspect ratio  $a/b = 0.33, 0.5, 1.0, 2.0, 3.0$; aspect ratios $a/b > 1$ correspond, as usual, to prolate spheroids. For all aspects ratios the volume of a single hard grain was kept the same (and equal to the volume of the approximately spherical grains with $D = 25$ nm). Volume concentration of the hard phase $c_{\rm hard} = 40\%$ was the same, as for simulations reported in the previous Sec. \ref{ExchWeakEffect}. The exchange weakening parameter $\kappa = 0.1$ was chosen close to the optimal value for spherical hard grains obtained above.

\subsubsection{Grain shape effect for $\mathrm{SrFe_{12}O_{19}/Fe}$}
\label{GrainShapeEffect_SrFe}

\begin{figure}[h!]
\centering
\resizebox{0.95\columnwidth}{!}{\includegraphics{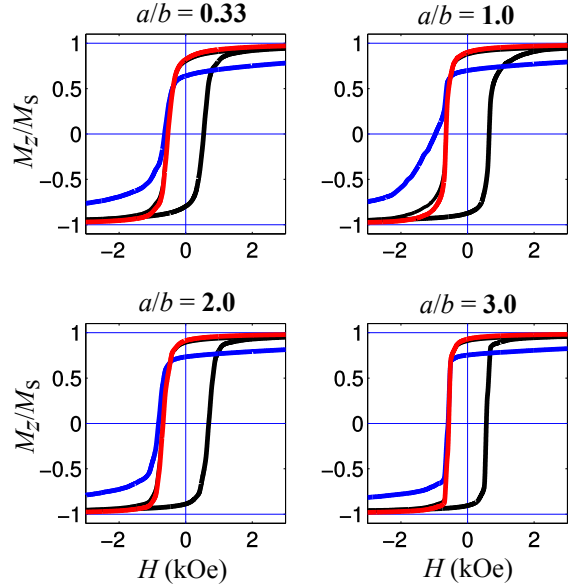}}
\caption{(color online). Simulated hysteresis curves of the nanocomposite $\mathrm{SrFe_{12}O_{19}/Fe}$ for the exchange weakening $\kappa=0.1$ and different aspect ratios of hard crystallites as indicated on the panels. Black loops - hysteresis of the total system, blue curves - upper part of the hysteresis loop for the hard phase, red curve - the same for the soft phase.}
\label{fighystellFe}
\end{figure}

\begin{figure}[h!]
\centering
\resizebox{0.95\columnwidth}{!}{\includegraphics{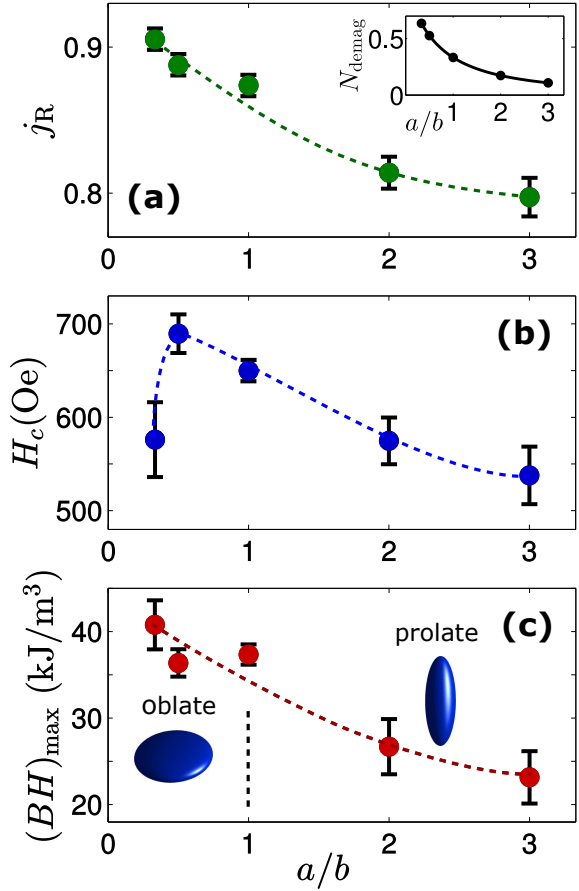}}
\caption{(color online). {\bf (a)} Simulated reduced remanence, {\bf (b)} coercivity and {\bf (c)} energy product of nanocomposites $\mathrm{SrFe_{12}O_{19}/Fe}$ with different aspect ratios $(a/b)$ of hard grains. Inset in (a) shows the demagnetizing factor in dependence on $a/b$. Dashed lines are guides for an eye.}
\label{figFeBHmax}
\end{figure}

\begin{figure}[h!]
\centering
\resizebox{0.95\columnwidth}{!}{\includegraphics{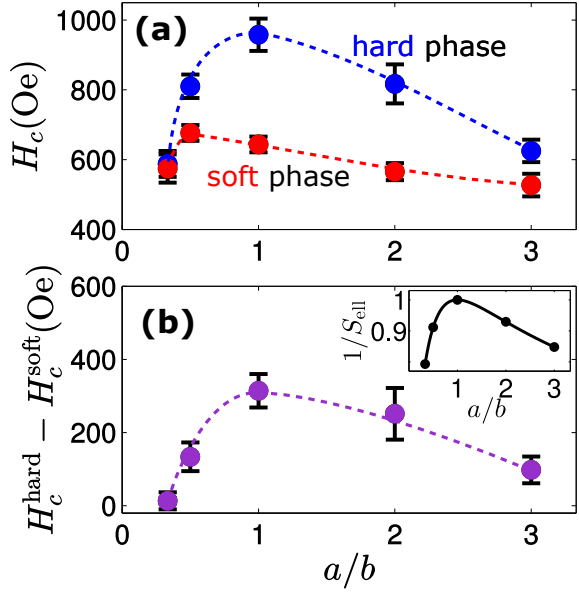}}
\caption{(color online) {\bf (a)} Coercivities of the hard (blue circles) $H_c^{\rm hard}$ and soft (red circles) $H_c^{\rm soft}$ phases and {\bf (b)} difference $\Delta H_c$  between these coercivities as functions of the aspect ratio $a/b$; inset in (b) - inverse of the surface area of an ellipsoid of revolution in dependence on $a/b$ . Dashed lines are guides for an eye. See text for the detailed explanation.}
\label{figFe_Hc_hs}
\end{figure}

First we discuss simulation results obtained for the composite $\rm{SrFe_{12}O_{19}/Fe}$ - see Figs. \ref{fighystellFe}, \ref{figFeBHmax} and \ref{figFe_Hc_hs}. In Fig.\ref{fighystellFe}, magnetization reversal curves for different aspect ratios $a/b$ are shown; both the loops for the entire system and for the soft and hard phases separately are presented. The most interesting observation here is the pronounced difference between the reversal curves of 'soft' and 'hard' phases for $a/b = 1$ and nearly synchronous magnetization reversal of both phases for other aspect ratios shown in the figure. This is a key feature for the understanding of the system behavior and will be discussed in detail below.

Overall dependencies of basic hysteresis parameters $j_R$, $H_c$ and $(BH)_{\rm max}$ on the aspect ratio $a/b$ is presented in Fig. \ref{figellFe111kappa}. Both main parameters of the hysteresis - remanence $j_R$ and coercivity $H_c$ exhibit a highly non-trivial dependence on this aspect ratio, which should be carefully analyzed.

The dependence $j_R(a/b)$ shown in Fig. \ref{figFeBHmax} is clearly counter-intuitive, because normally one would expect a {\it higher} remanence for a system containing elongated particles - in our case for $a/b > 1$ - due to the positive shape anisotropy constant for such particles. The simulated dependence shows the opposite trend - the remanence increases with decreasing the aspect ration $a/b$, i.e., $j_R$ becomes larger for a composite with {\it oblate} hard grains.

This behavior can be explained taking into account that hard ellipsoidal grains are mostly embedded into the soft magnetic matrix (soft phase), which magnetization is larger than that of the hard phase: $M_{\rm Fe} > M_{\rm SrFe_{12}O_{19}}$. This means that hard grains represent magnetic 'holes' inside a soft matrix, what means, in turn, that the total magnetodipolar field acting on the magnetization of the hard grain, is directed (on average) towards the initially applied field. With another words, this field acts as a {\it magnetizing} field, i.e. it increases the remanence of the hard phase. 

The magnitude of this magnetizing field is proportional to the difference between magnetizations of the soft and hard phases and is of the order $H_{\rm dip}^{\rm mag} \sim N_{\rm dem} \cdot (M_{\rm Fe} - M_{\rm SrFe_{12}O_{19}}) = N_{\rm dem} \cdot \Delta M$. For our system parameters $\Delta M = 1300 \, {\rm G}$, so that, taking into account that $N_{\rm dem}  \sim \pi$, we obtain $H_{\rm dip}^{\rm mag} \sim 4 \, {\rm kOe}$. This value is comparable to the mc-anisotropy field of the hard grain itself ($H_K({\rm SrFe_{12}O_{19}}) = 20 \, {\rm kOe})$, so the effect of this magnetodipolar field can be significant.

To explain the trend $j_R(a/b)$ seen in Fig. \ref{figFe_Hc_hs}, it remains only to note that this magnetizing field is larger for {\it oblate} spheroids, for which it can achieve the magnitude of $4\pi \Delta M_s$ - the limiting case for a thin disk with the revolution axes along the magnetizing direction of the system. In contrast, for the prolate spheroid $H_{\rm dip}^{\rm mag}$ becomes weaker when $a/b$ increases (spheroid becomes more prolate), because the main contribution to this field comes from the soft phase regions near the ends of this prolate spheroid. 

The result of this complicated interplay is the better alignment of magnetic moments of the hard phase consisting of oblate particles. This leads to the higher remanence of the whole system for two reasons: ({\it i}) the remanence of the hard phase itself is larger and ({\it ii}) the 'supporting' action of the hard phase on the soft phase - due to the interphase exchange coupling - is more significant.

The explanation of the non-trivial dependence of the coercivity on the aspect ration $H_c(a/b)$ - with the maximum between $a/b =0.5$ and $a/b =1.0$ - requires a detailed understanding of the magnetization reversal mechanism in composites with partial interphase exchange coupling.

Namely, magnetization reversal of these nanocomposites always occurs according to the following scenario: the soft phase switches first, and then exhibit a torque on the hard grains due to the interphase exchange interaction. For non-negligible interphase exchange this torque is the main interaction mechanism between the phases and leads (together with the applied field) to the magnetization reversal of the hard phase in larger negative external fields.

In order to understand, why the coercivity has its maximum for particles with a weak shape anisotropy, we have to recall that the interphase exchange interaction is a surface effect and as such is proportional to the interphase surface area. In our case this is the surface area of hard grains, which are mostly surrounded by the soft phase. This means, that exchange torque which the soft phase exhibits on the hard grains, is proportional to the surface area of these grains. Hence, this torque should be minimal for the hard grains with the spherical shape, because the surface area of an ellipsoid of revolution with the given volume is minimal for $a/b = 1$ (sphere).

For this reason hard phase with grains having the shape close to spherical will have the maximal coercivity, i.e. reverse in the largest negative field. Such grains will be also able to 'support' soft phase up to negative fields larger than non-spherical hard grains would do, leading to the largest coercivity of the whole sample.

To provide further proof of this hypothesis, we have plotted in Fig. \ref{figFe_Hc_hs} the coercivities of the hard and soft phases separately (see curves for $H_c^{\rm hard}$ and $H_c^{\rm soft}$ on the panel (a)) and the difference between them $\Delta H_c$ on the panel (b) as functions of the aspect ratio $a/b$. The excellent qualitative agreement between $\Delta H_c(a/b)$ and the inverse of the surface area of an ellipsoid of revolution $1/S_{\rm ell}(a/b)$ (see inset to this panel) as the functions of $a/b$ clearly shows that the observed effect is due to the surface-mediated interaction, what in our case clearly means the interphase exchange interaction.

We finish this subsection with the explanation why the dependence of the maximal energy product on the aspect ratio $E_{\rm max}(a/b)$ (panel (c) in Fig.\ref{figFeBHmax}) for our system closely follows the corresponding trend of the remanence $j_R(a/b)$ (see panel (a)), but is not influenced by the dependence $H_c(a/b)$ (panel (b)).

To understand this phenomenon, we recall that the energy product is defined as the maximal value of the product $(BH)$ within the second quadrant of the hysteresis loop, i.e. for external fields $-H_c < H < 0$ (here and below we omit for simplicity the index $z$ by $H$, $B$ and $M$):

\begin{equation}
\label{EnProdDef}
E_{\rm max} = \max_{-H_c < H < 0}[B(H) \cdot H] = \max[(H + 4\pi M(H)) \cdot H],
\end{equation}
where it is important, that the energy product depends on $H$ both explicitly and implicitly - via the dependence $M(H)$.

Let us now assume that for some reference parameter value (e.g. in our case for $a/b=1$) with the magnetization vs. field dependence given by the function $M_{\rm ref} (H)$ the product (\ref{EnProdDef}) reaches its maximum $E_{\rm max}^{(0)}$ for the field value $H_0$. This means that the corresponding derivative of the energy product $dE/dH$ vanishes at this point, leading to the condition

\begin{equation}
\label{DerivEnProd}
\begin{split}
{\frac{d}{dH}}[(H + 4\pi M_{ref}(H)) \cdot H]|_{H = H_0} =  \\ 
= H_0 + 2\pi \left[ M_{\rm ref}(H_0) + H_0 \frac{dM_{\rm ref}}{dH} \bigg|_{H = H_0} \right] = 0
\end{split}
\end{equation}

If the parameter in question changes (i.e. we take another value of $a/b$), then the hysteresis loop also changes, becoming $M_{\rm new} (H) = M_{\rm ref} (H) + \Delta M$ and the maximum of the energy product is achieved at another field $H_{\rm new} = H_0 + \Delta H$. The new maximal energy product then is

\begin{equation}
\label{NewEnProd}
E_{\rm max}^{\rm new} = (H_{\rm new} + 4\pi M_{\rm new}(H_{\rm new})) \cdot H_{\rm new}]. 
\end{equation}

Assuming that $\Delta M$ and $\Delta H$ are small, we can expand the functions $M_{\rm ref}(H)$ and $\Delta M(H)$ in the vicinity of the point $H_0$. Starting from Eq.\ref{NewEnProd} and retaining only terms linear in $\Delta M$ and $\Delta H$, we obtain for $E_{\rm max}^{\rm new}$ the expression

\begin{equation}
\label{NewEnProdExpand}
\begin{split}
E_{\rm max}^{\rm new} = E_{\rm max}^{(0)} + 4\pi H_0 \Delta M(H_0) + \\
+ 2 \Delta H 
\left\lbrace
H_0 + 2\pi \left[ M_{\rm ref}(H_0) + H_0 \frac{dM_{\rm ref}}{dH} \bigg|_{H = H_0} \right]
\right\rbrace
\end{split}
\end{equation}

The coefficient in braces after $\Delta H$ is exactly the expression (\ref{DerivEnProd}) at the point where the reference energy product reaches its maximum and as such is equal to zero. Hence we are left with the expression of the new maximal energy product as

\begin{equation}
\label{NewEnProdFinal}
E_{\rm max}^{\rm new} = E_{\rm max}^{(0)} + 4\pi H_0 \Delta M(H_0)
\end{equation}

This expression shows that the change of the energy product due to the variation of some external system parameter is controlled mainly by the change of the magnetization curve (upper part of the $M-H$ hysteresis loop) at the external field $H_0$ where the reference energy product reached its maximum. Obviously, this magnetization change is roughly proportional to the remanence change, which explains the semiquantitative correspondence between the dependencies $j_R(a/b)$ and $E_{\rm max}(a/b)$. We point out once more that this statement is true only if the coercivity shift due to the variation of this parameter is relatively small. If this is not the case, then the position of the vertical side of the rectangle (in the $(B-H)$ plane) used for the determination of $(BH)_{\rm max}$ can strongly depend on the coercivity value, making the derivation of Eq. (\ref{NewEnProdExpand}) invalid. 

\subsubsection{Grain shape effect for $\mathrm{SrFe_{12}O_{19}/Ni}$}
\label{GrainShapeEffect_SrNi}

The second composite which we have used to study the grain shape effect - ${\rm SrFe_{12}O_{19}/Ni}$ - is qualitatively different from the previous material (${\rm SrFe_{12}O_{19}/Fe}$) due to the much lower magnetization of the soft phase: $M_s({\rm Ni}) \approx 490 \, \rm{G}$. The idea behind the usage of a soft phase with such a low magnetization is that the coercivity of the resulting material should be much higher due to the weaker response of the soft phase with a smaller magnetization to the external field. This higher coercivity might compensate the decrease of the net magnetization, resulting in a competitive energy product.

\begin{figure}[h!]
\centering
\resizebox{0.95\columnwidth}{!}{\includegraphics{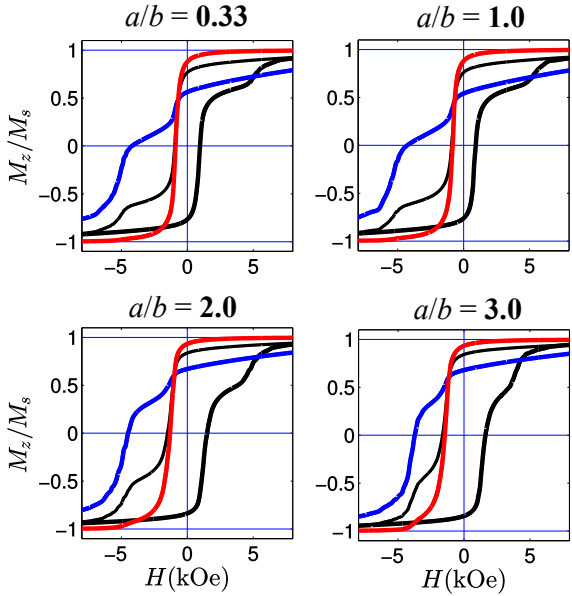}}
\caption{(color online).Simulated hysteresis curves of the composite $\mathrm{SrFe_{12}O_{19}/Ni}$ for $\kappa=0.1$ and different aspect ratios of hard crystallites as indicated on the panels. Black loops - total hysteresis, blue curves - upper part of the hysteresis loop for the hard phase, red curve - for the soft phase.}
\label{fighystellNi}
\end{figure}

\begin{figure}[h!]
\centering
\resizebox{0.95\columnwidth}{!}{\includegraphics{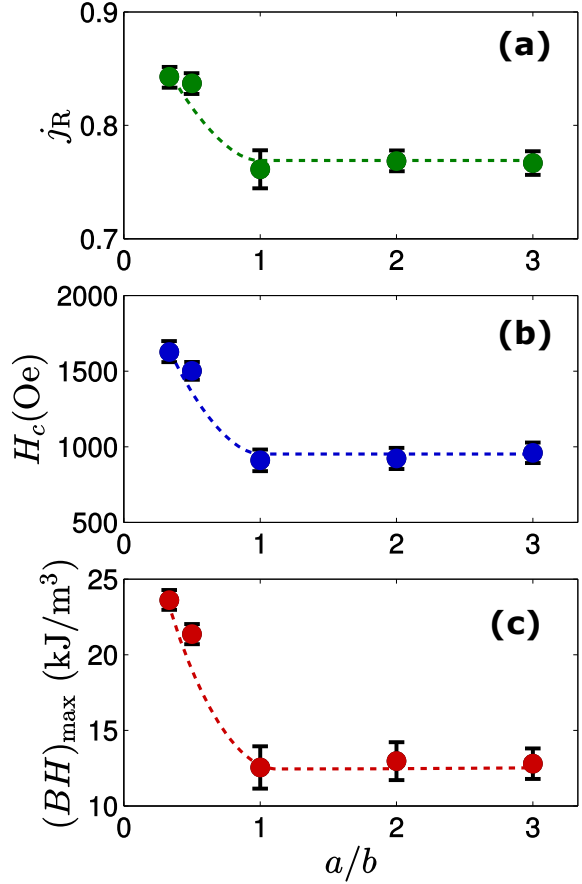}}
\caption{(color online). {\bf (a)} Remanence, {\bf (b)} coercivity and {\bf (c)} energy product of the nanocomposite $\mathrm{SrFe_{12}O_{19}/Ni}$ in dependence on the aspect ratio $a/b$ of hard grains. Dashed lines are guides for an eye.}
\label{figNiBHmax}
\end{figure}

Simulation results for $\rm{SrFe_{12}O_{19}/Ni}$ with various grain shapes are presented in Fig. \ref{fighystellNi} (hysteresis loops) and Fig. \ref{figNiBHmax} (basic characteristics of the hysteresis). As it can be clearly seen, the Ni-containing composite behaves itself qualitatively different compared to Fe-containing material.

The main new feature is - as expected - the higher coercivity of both soft and hard phases. The considerably larger coercivity of the Ni phase ($H_c({\rm Ni}) \approx 1000 \, \rm{Oe}$ - see Fig. \ref{fighystellNi}, red loops) compared to the coercivity of the Fe phase in the previously studied composite ($H_c({\rm Fe}) \approx 600 \, \rm{Oe}$ - see Fig. \ref{figFe_Hc_hs}a) is mainly due to the lower magnetization of Ni, as mentioned above. 

The much larger coercivity of the hard phase (we remind that this phase is the same for both studied materials) for  $\rm{SrFe_{12}O_{19}/Ni}$ can be most probably explained by three reasons. First, due to the lower $M_s(\rm{Ni})$ the reversal of the soft phase starts for the Ni-containing composite in higher negative fields, what should by itself lead to larger  $H_c^{\rm hard}$ for the hard phase also. However, this reason alone could not be responsible for the more than fourfold increase of $H_c^{\rm hard}$ ($H_c^{\rm hard} (\rm{SrFe_{12}O_{19}/Ni}) \approx 4000 \, \rm{Oe}$ vs  $H_c^{\rm hard}(\rm{SrFe_{12}O_{19}/Fe}) < 1000 \, \rm{Oe}$). 

The second reason is that the magnetodipolar field of the soft phase acting on hard grains is much smaller for Ni- than for Fe-containing composites (due to the same much lower magnetization of Ni). After the reversal of the soft phase this magnetodipolar field is directed opposite to the initial material saturation and thus assists the reversal of hard grains. Much smaller magnitude of this field thus leads to much higher external field required to reverse the hard phase.

Finally, the considerably smaller exchange constant of Ni compared to Fe (see Table 1) results in the lower exchange torque acting on hard grains after the reversal of the soft phase, also decreasing the total torque acting on the hard phase and increasing its coercivity.

This qualitatively new situation - the non-correlated magnetization reversals of the soft and hard phases - leads to the another type of the coercivity dependence on the aspect ratio $a/b$ of hard grains: coercivity $H_c (a/b)$ decreases when this ratio increases (see Fig. \ref{figNiBHmax}b), behaving itself in a very similar way as the remanence $j_R(a/b)$ (Fig. \ref{figNiBHmax}a). 

The most likely explanation of this behavior is the following: the degree of the magnetization alignment of hard grains for their different aspect ratios is nearly the same for various values of $a/b$. Hence the main effect on the soft phase reversal is due to the difference in the magnetodipolar field distributions caused by hard grains having various shapes. Prolate ellipsoids of revolution produce a non-uniform magnetodipolar field which maximal value is higher than for oblate ellipsoids. However, the field of prolate particles is strongly concentrated near their 'sharp' ends, whereas the dipolar field of oblate ellipsoids (magnetized on average along their axes of revolution) occupies a much larger region near their 'flat' surfaces. For this reason  the dipolar field of oblate hard grains supports the magnetization of the soft phase in larger space regions, thus leading to the increase of the soft phase (and total) coercivity with decreasing $a/b$, i.e. when hard grains become more oblate. 

As the result of the decrease of both $j_R$ and $H_c$, the energy product also decreases with increasing $a/b$, as shown in Fig. \ref{figNiBHmax}c. We point out that although the resulting behavior is qualitatively somewhat similar to the case of the Fe-containing material (compare to Fig. \ref{figFeBHmax}) - the energy product decreases with increasing aspect ratio - the physical reasons for this behavior are fundamentally different for these two composites.

It also interesting to note that despite the larger coercivity of Ni-containing composite, its energy product remains smaller than for the Fe-containing material due to its much lower net magnetization. This relation can change for materials with different fractions of soft and hard phases.

\section{Conclusion}
\label{Conclusion}

In this paper we have presented a detailed numerical study for the dependence of magnetic properties of Sr-ferrite-based nanocomposites on two very important material parameters: ({\it i}) exchange coupling between various crystallites (what includes the coupling between soft and hard grains) and ({\it ii}) the shape of the hard grains.

First, we have demonstrated - in contrast to the common paradigm - that the maximal energy product $E_{\max} = (BH)_{\rm max}$ is a non-monotonous function of the intergrain exchange coupling $\kappa$ and that the optimal $\kappa$-value  (for which the energy product reaches its highest value) is far below the perfect coupling. This non-monotonous character of the function  $E_{\max}(\kappa)$ is due to the corresponding dependence of the coercivity on the exchange coupling $H_c(\kappa)$. 

Second, we have studied the dependence of the hysteresis properties and the maximal energy product on the shape of hard grains for two very different nanocomposite materials - $\rm{SrFe_{12}O_{19}/Fe}$ ($M_s({\rm Fe}) \approx 1700 \, {\rm G}$) and $\rm{SrFe_{12}O_{19}/Ni}$ ($M_s({\rm Ni}) \approx 490 \, {\rm G}$). Hard grains have been assumed to have (approximately) a shape of ellipsoids of revolution, which aspect ratio was varied in the range $1/3 \leq a/b \leq 3$ ($a/b > 1$ corresponds to a prolate ellipsoid). We have shown, that for both materials the aspect ratio dependence of the maximal energy product $E_{\max}(a/b)$ essentially follows the corresponding dependence of the hysteresis loop remanence $j_R(a/b)$ and have supported this observation by analytical considerations. For both materials, the maximal value of $j_R(a/b)$ - and hence of $E_{\max}(a/b)$ - was obtained for the {\it oblate} hard grains with the smallest aspect ration $a/b = 1/3$ (also in contrast with common expectations). Physical reasons for this behavior are revealed.

Finally, we have also analyzed the dependence of the coercivity on the shape of hard grains $H_c(a/b)$ and have shown that this dependence for the two composites under study is qualitatively different. For $\rm{SrFe_{12}O_{19}/Fe}$ the function $H_c(a/b)$ has a pronounced maximum for approximately spherical grains, whereas for $\rm{SrFe_{12}O_{19}/Ni}$ coercivity monotonously decreases with increasing $a/b$. This difference is explained by analyzing the dominating interaction mechanisms between the hard and soft phases in these materials.

\section{Acknowledgment}
\label{Acknowledgment}
Financial support of the DFG project BE2464/10-3 and the EU-FP7 project "NANOPYME" (310516) [www.nanopyme-project.eu] is greatly acknowledged. Also we would like to thank Dr. Alberto Bollero, Dr. Adri\'an Quesada and Dr. Cesar de Julian Fernandez for fruitful discussions and Dr. Andreas Michels for critical reading of the manuscript.

\bibliography{nanocomp_mumag}
\bibliographystyle{apsrev4-1}

\end{document}